\documentclass[%
 reprint,
 amsmath,amssymb,
 aps,
]{revtex4-1}

\usepackage{graphicx}% Include figure files
\usepackage{dcolumn}% Align table columns on decimal point
\usepackage{bm}% bold math
\usepackage{cancel}
\usepackage{amsmath}
\usepackage{hyperref}% add hypertext capabilities
\usepackage{empheq}
\usepackage{multirow}
\usepackage{cleveref}
\usepackage{xcolor}
\usepackage[mathlines]{lineno}% Enable numbering of text and display math

\def\c{{\rm c}}

\begin{document}

\preprint{Prepared for submission to PRD}

\title{Tracing the anomalous $tqg$ and $tq\gamma$ flavor changing interactions at the FCC-he}% Force line breaks with \\

\author{Subhasish Behera}
\email{subhparasara@gmail.com}
\altaffiliation[Also at ]{Institute of Physics, Johannes Gutenberg University Mainz, Germany.}%Lines break automatically or can be forced with \\

\author{Poulose Poulose}%
\email{poulose@iitg.ac.in}
\affiliation{Department of Physics, Indian Institute of Technology Guwahati, India.}%


\begin{abstract}
	We investigate the possible presence of Flavor Changing Neutral Current (FCNC) couplings of the top quark with gluon and photon through $e^-p\to e^-tj$ process at the Future Circular Collider in the proton-electron mode (FCC-he). Focusing on disentangling the effects of different couplings that could be present, we exploit the presence of the scattered electron, the angular distribution of which is sensitive to the type of coupling involved. Top quark polarisation accessed through the angular distribution of the decay lepton provides additional handle in identifying the nature of the couplings. Further, we demonstrate the potential of electron beam polarisation in distinguishing the left-handed and right-handed couplings of both gluon and photon separately. Considering an $e^-p$ collider of beam energies of $E_{e(p)} = 60~(50000)$~GeV at 2~ab$^{-1}$ integrated luminosity, couplings can be probed at the level of $10^{-2}$ with the corresponding branching fractions of ${\rm BR}(t\to u\gamma)\le 4 - 7 \times 10^{-6}$ and ${\rm BR}(t\to c\gamma) \le 1-2 \times 10^{-5}$, depending on if the coupling is right-handed or left-handed. The corresponding limits on the gluon couplings lead to ${\rm BR}(t\to ug)\le 1.7 \times 10^{-5}$ and ${\rm BR}(t\to cg) \le 3-4 \times 10^{-5}$.
\end{abstract}

\pacs{FCC-he, Top quark, FCNC}%Use showkeys class option if keyword
\maketitle

%\tableofcontents

\section{Introduction}

Top quark has a special place in elementary particle dynamics. Discovered a little more than two decades ago, this heaviest elementary particle about 185 times more massive than a proton %weighing about 185 times that of a proton 
is the only quark that decays weakly as a bare quark, the hadronization time being larger than its mean life time. This provides a unique window to explore the weak interactions of the quark sector in a direct way. Within the Standard Model (SM), quarks can have both charged current interactions with the mediation of charged gauge bosons, as well as neutral current interactions mediated by neutral gauge bosons. While by nature charged current interactions couple quarks of different flavours, Flavour Changing Neutral Current (FCNC) interactions are forbidden due to the GIM Mechanism \cite{Glashow:1970gm}. Thus, while $t\to Wb$ decay is the most favoured channel, $t\to Zc,~\gamma c,~gc$ (where $g$ represents a gluon) or those with $c$ quark replaced by $u$ quark are extremely rare. Absent at tree level in the SM, higher order quantum corrections lead to ${\rm BR}(t\to (Z/\gamma/g)c) \sim 10^{-14}$, which is suppressed by another two to three orders of magnitude in the case of $u$ quarks \cite{AguilarSaavedra:2004wm}. 
%{Agashe:2013hma}
The experimental measurements, on the other hand, are not yet capable of reaching out to such precision.
Coming to the present constraints, 
the ATLAS collaboration of the LHC has performed a search for  $t\gamma$ events with  81~fb$^{-1}$ data at $\sqrt{s}=13$~TeV~\cite{Aad:2019pxo}.
%Sirunyan:2017uae, CMS:2013nea}.
Subsequently, they placed  95\% C.L.  bounds of  ${\rm BR}(t\to \gamma u) \leq 2.8\times 10^{-5}$ for left-handed couplings and $6.2\times 10^{-5}$ for right-handed couplings;  and  ${\rm BR}(t\to \gamma c) \leq 22\times 10^{-5}$  and $88\times 10^{-5}$ for the left- and right-handed couplings, respectively.
This is about an order of magnitude improvement on the bounds coming from a similar search by the CMS experiments \cite{Khachatryan:2015att}.
Coming to $tqg$ couplings, ATLAS collaboration searching for $qg\to t\to Wb$ process sets a limit on the branching fractions,
${\rm BR}(t\to g u) \leq 4.5\times 10^{-5}$ and  ${\rm BR}(t\to g c) \leq 20 \times 10^{-5}$ using the 20.3 fb$^{-1}$ data collected at the centre of mass energy of 8 TeV \cite{Aad:2015gea} . CMS search limits these coupling to $2.0\times 10^{-5}$ and $41\times 10^{-5}$, respectively, from the 5 and 19.5 fb$^{-1}$ data collected at $\sqrt{s}=7$ and 8 TeV \cite{Khachatryan:2016sib}. Anomalous top-$Z$ FCNC couplings are bound by the ATLAS search for rare decays of the top quark in $t\bar t$ events at $\sqrt{s}=13$ TeV with a data set of 36.1 fb$^{-1}$, setting a limit of ${\rm BR}(t\to Z u) \leq 1.7\times 10^{-4}$ and ${\rm BR}(t\to Z c) \leq 2.4\times 10^{-4}$ \cite{Aaboud:2018nyl, ATLAS:2019pcn}. 
Top-Higgs anomalous FCNC couplings are limited by the CMS search with $pp\to tH$ at centre of mass energy of 13 TeV with a data set of 35.9 fb$^{-1}$, bounding 
${\rm BR}(t\to H q) \leq 0.47\%$, for both $q=u,~c$ \cite{Sirunyan:2017uae}. The ATLAS bounds on $tqH$ FCNC from $tH$ production at 13 TeV LHC with a data of 36.7 fb$^{-1}$ is ${\rm BR}(t\to H u) \leq 2.4 \times 10^{-3}$ and ${\rm BR}(t\to H c) \leq 2.2\times 10^{-3}$ \cite{Aaboud:2017mfd}.
At the same time, projected reach of these {\rm BR}'s at the high luminosity LHC with 3 ab$^{-1}$ luminosity (HL-LHC) are  $2.5-5.5 \times 10^{-5}$ \cite{ATL-PHYS-PUB-2016-019}, which is at the best an order of magnitude better than the current measurements. With the standard production of both the single top as well top-antitop pair in plenty, it is hard to probe anomalous top quark couplings in production at the LHC. A phenomenological study for top-antitop quark production at the Future Circular Collider in the proton-proton mode (FCC-hh) with 100 TeV center of mass at $\mathcal{L}$ of 10 ab$^{-1}$ sets limit on  the {\rm BR}($t\to Hc$) $\le$ {$\mathcal{O} (10^{-3})\%$} \cite{Papaefstathiou:2017xuv}. On the other hand, triple top quark production at the high energy versions of proton-proton collisions (HE-LHC and FCC-hh) has been proposed to probe the presence of FCNC, as the standard production in this case is forbidden \cite{Cao:2019qrb, Khanpour:2019qnw, Oyulmaz:2018irs}. It may be noted that colliders with electron beam has multiple advantages in probing the top quark FCNC, as illustrated in Ref. \cite{Cakir:2018ruj, Behera:2018ryv, Liu:2019wmi, Alici:2019asv, Oyulmaz:2019jqr}.  In particular,  in Ref. \cite{Behera:2018ryv} we have pointed out the possibility of distinguishing different Lorentz structures of the coupling by studying the angular distributions of scattered electron and the asymmetries associated with it. In a study of %Studying 
the single top production at FCC-ee, Ref. \cite{Khanpour:2014xla} has shown that the $Ztq$ and $\gamma tq$ couplings can be probed with the corresponding {\rm BR} to the level of $3-5 \times 10^{-5}$ at a center of mass energy of 350 GeV with moderate luminosity of 300~fb$^{-1}$.

\begin{figure}[tph]
	\centering
	\includegraphics[trim=0 0 0 0,clip,width=0.20\linewidth]{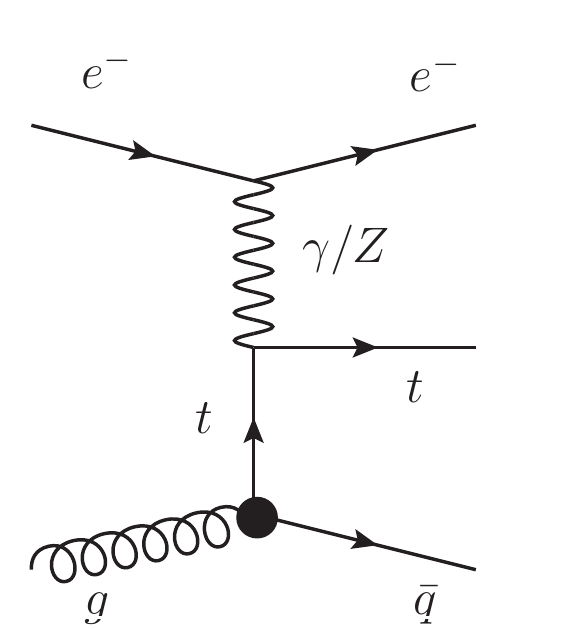}
	\includegraphics[trim=0 0 0 0,clip,width=0.20\linewidth]{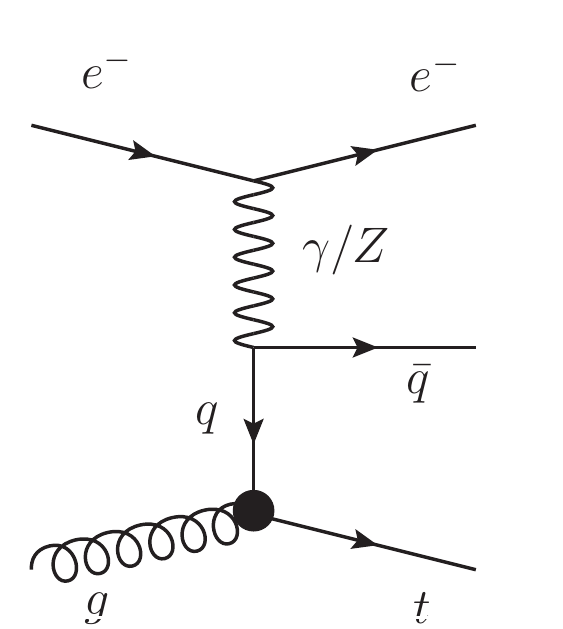}
	\includegraphics[trim=0 0 0 0,clip,width=0.20\linewidth]{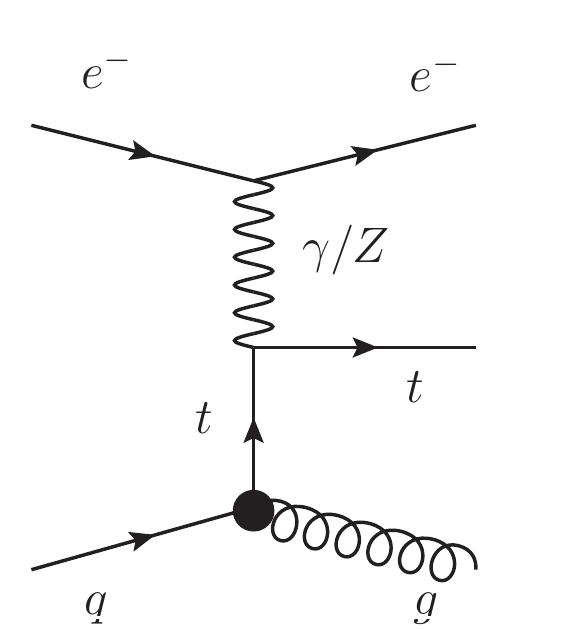}
	\includegraphics[trim=0 0 0 0,clip,width=0.25\linewidth]{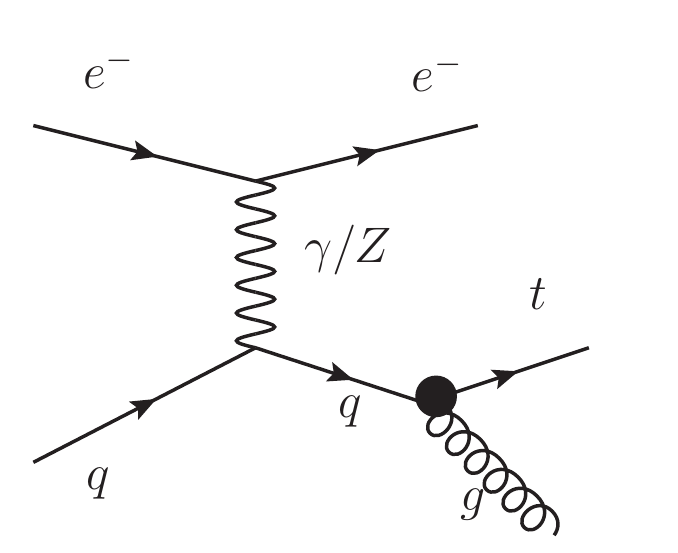}
	\\
	\includegraphics[trim=0 0 0 0,clip,width=0.20\linewidth]{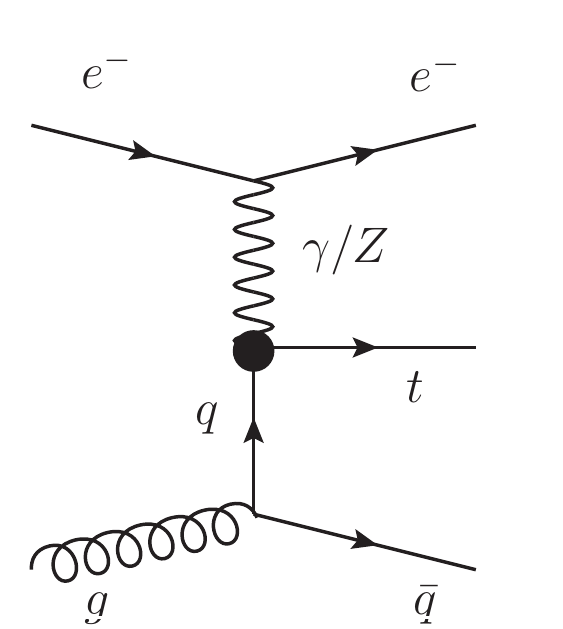}
	\includegraphics[trim=0 0 0 0,clip,width=0.20\linewidth]{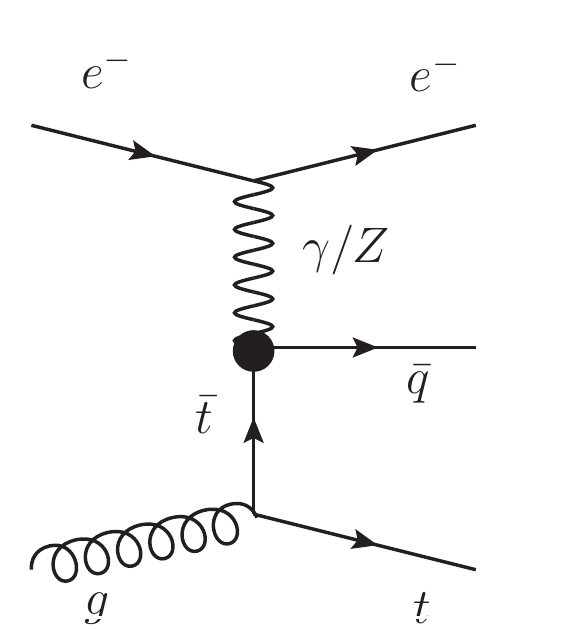}
	\includegraphics[trim=0 0 0 0,clip,width=0.20\linewidth]{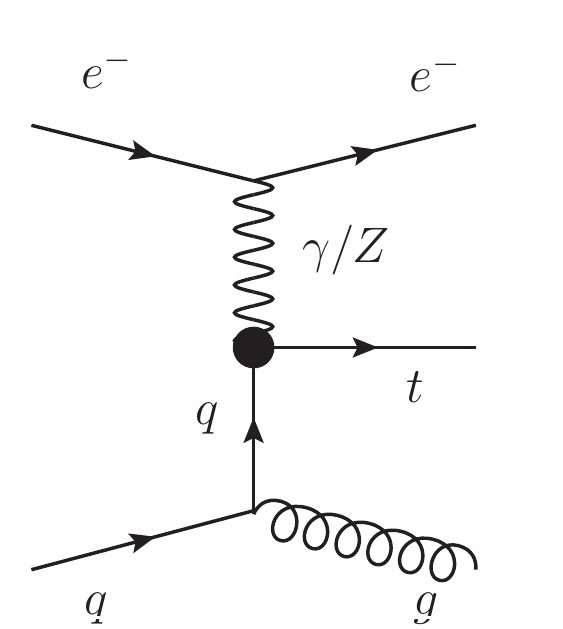}
	\includegraphics[trim=0 0 0 0,clip,width=0.25\linewidth]{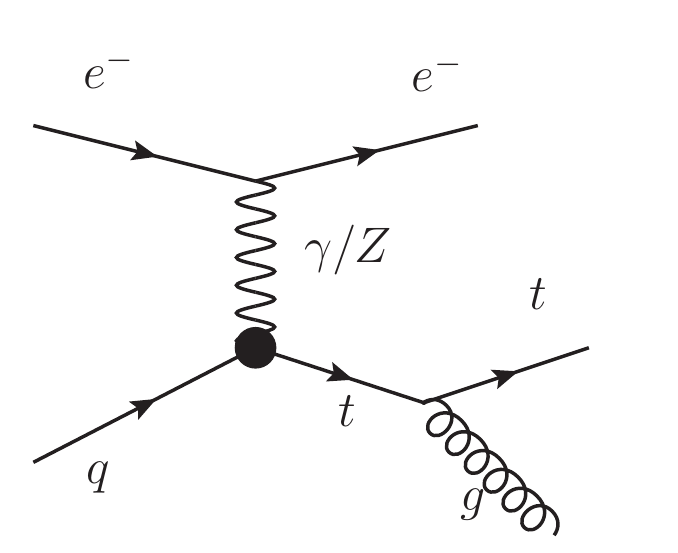}
	\caption{\small  Feynam diagrams representing the process $ep\to e t j$, where $j$ is either a light quark or a gluon. In this study we consider $q=u,~c$.}    \label{fig:EW_FCNC_signal}
\end{figure}

In this report, we shall study the process $ep\to ejt$, with subsequent leptonic decay of the top quark. We shall show that this process is suitable to probe $tqg$ coupling in an effective way along with the $tq\gamma$ coupling. We parametrise the FCNC interactions of the top quark through the effective Lagrangian  \cite{AguilarSaavedra:2008zc},

\begin{align}\label{eq:Lag}\nonumber
- {\cal L}_{\rm fcnc}
=& g_s\bar{q}\lambda^a\frac{i \sigma^{\mu \nu}q_{\nu}}{\Lambda}( \kappa_{gqt}^L P_L +  \kappa_{gqt}^R P_R)tG_{\mu}^a \\\nonumber&
+e\bar{q}\frac{i \sigma^{\mu \nu}q_{\nu}}{\Lambda}( \kappa_{\gamma qt}^L P_L +  \kappa_{\gamma qt}^R P_R)tA_{\mu}  \\\nonumber&
+ \frac{g}{2c_W}\bar{q}\gamma^{\mu}( X_{zqt}^L P_L +  X_{zqt}^R P_R)tZ_{\mu} \\&
+ \frac{g}{2c_W}\bar{q}\frac{i \sigma^{\mu \nu}q_{\nu}}{\Lambda}( \kappa_{zqt}^L P_L +  \kappa_{zqt}^R P_R)tZ_{\mu} + {\rm H.c},
\end{align}
where $q = u,~c$ quark, $q_\nu=p_t - p_q$, is the momentum transfer between the top quark with momentum $p_t$ and the light quark with momentum $p_q$ in the process, and $\Lambda$ is the cut-off scale. The vector couplings are denoted by $X_{zqt}^{L,R}$ for $Z$-boson and the tensor couplings by $\kappa_{gqt}^{L,R}$, $\kappa_{\gamma qt}^{L,R}$ and $\kappa_{zqt}^{L,R}$ for gluon, photon and $Z$-boson, respectively. Other symbols have their usual meaning. The Feynman diagrams corresponding to $ep\to etj$ in the presence of these anomalous couplings are illustrated in Fig.~\ref{fig:EW_FCNC_signal}.  
The scattered electron, posing like an innocent spectator turns out to be a valuable informer capable of providing clear indication of the Lorentz structure of the $tqg(\gamma)$ couplings. This possibility alone is a marked advantage of the $ep$ collider against the $pp$ colliders. Demonstrating this, we shall study the angular asymmetry of the scattered electron as a useful discriminator between the gluon and photon couplings.  

We shall organize this report as follows. In \cref{sigbkg} we discuss the signal and background process in details with cross section for different polarization of initial electron beam. In \cref{eventgen} we discuss the event generation and analysis. In \cref{ch_5:spin_matrix} we discuss various asymmetries and explore the possibility of discriminating different couplings. In \cref{ch_5conclusion} we summarise the study and present our conclusions.

\section{Signal and Background} \label{sigbkg}
There are two major deep-inelastic scattering facilities proposed by CERN, namely the Large Hadron electron Collider (LHeC) \cite{AbelleiraFernandez:2012cc} and the Future Circular Collider in the proton-electron mode (FCC-he) \cite{Abada:2019lih}. LHeC intend to use the same tunnel as that of the presently running LHC, replacing one of the proton beams with an electron beam. The proposed baseline energy of the electron beam is 60 GeV, which in combination with 7 TeV proton beam results in a center of mass energy of 1.3 TeV. This is about four times that of its predecessor, HERA at DESY with centre of mass energy of 319 GeV. In addition the luminosity expected is about two orders of magnitude higher. The case of FCC is more ambitious with a baseline proton energy of 50 TeV, keeping the electron energy of 60 GeV, corresponding to a center of mass energy of 3.5 TeV. Luminosity of FCC-he is expected to be about an order larger than that of the LHeC. We refer the reader to the CDR for the details of physics case of these proposed projects \cite{AbelleiraFernandez:2012cc, Abada:2019lih}.  Presence of electron beam in the initial state make it unique in probing the properties of particle dynamics, which is not possible in proton-proton collisions. We had demonstrated this in a previous work \cite{Behera:2018ryv}, where we studied the anomalous $tqZ$ couplings, illustrating the use of the scattered electron in disentangling the effect of different couplings involved, in the context of LHeC.  In the present case, we propose to study the $tgq$ and $t\gamma q$ couplings through single top quark production along with a light jet and the scattered electron. At LHeC with proton beam energy of 7 TeV,  the cross section for $p e^- \to e^- tj, (t \to Wb, W \to \ell \nu_\ell)$ process is about 0.05 fb and 0.1 fb in the case of gluons and photon couplings, respectively, with the value of the coupling taken as 0.01 TeV$^{-1}$. For enhanced cross section, we turn to the proposed FCC-he with possible proton beam energy of 50 TeV. The electron beam energy is set to the proposed baseline value of 60 GeV in our analysis.
At parton level the signal final state consists of $e^-b\ell\nu j$, where $\ell=e,~\mu$. This final state does not arise in the SM for hard scattering process in $ep$ collision. However, the SM processes with final states (i) $e^-jj\ell\nu$,  (ii) $e^-bjj\ell\nu$, (iii) $bjjj\nu$,  (iv) $bjj\nu$ and (v) $e^-jj \ell \nu$ could potentially mimic the signal after showering and hadronization. As the first step, we compute the signal and background cross section with the help of MadGraph5 \cite{Maltoni:2002qb, Alwall:2011uj}.
In \cref{tab:CH_5_sig_cs}, we have listed the cross section of the signal process for the production of top quark or anti-top quark with their subsequent leptonic ($e^\pm,~\mu^\pm$) decay, represented by $\sigma_{t}$ and $\sigma_{\bar{t}}$, respectively, for different beam polarisations. The values quoted are for the $u$ quark coupling with the top quark, which are slightly reduced when $c$ quark is considered (owing to the smaller \emph{Parton Distribution Function} (PDF)) as listed in \cref{tab:CH_5_sig_cs_cquark}. We  have considered one coupling to be present at a time setting its value to be unity, while setting all others to zero. Noticing that the cross section scales like the square of the coupling, one can obtain this for any value of the coupling in a straight forward way. If multiple couplings are considered, then there would be interference terms between those couplings in some scenarios.

\begin{table}[tph]
	\centering
	\resizebox{0.99\linewidth}{!}
	%\small
	{
		\begin{tabular}{c|r|r||r|r||r|r|r|r} \hline
			Coupling & $\kappa_{gut}^L$ & $\kappa_{gut}^R$ & $\kappa_{\gamma ut}^L$ & $\kappa_{\gamma ut}^R$ & $X_{zut}^L$  & $X_{zut}^R$ & $\kappa_{zut}^L$  & $\kappa_{zut}^R$  \\ \hline \hline
			\multicolumn{9}{c}{unpolarized} \\\hline\hline
			%$\sigma_{t}$ (fb)          &526.06&556.60&1601.32&1602.06&263.35&251.40&310.44&336.57 \\ \hline 
			$\sigma_{t}$ (fb)          &526&557&1601&1602&263&251&310&337 \\ \hline 
			% $\sigma_{\bar{t}}$ (fb) &259.03&264.84&835.46&832.39&141.06&147.79&122.79&106.24  \\ \hline \hline
			$\sigma_{\bar{t}}$ (fb) &259&265&835&832&141&148&123&106  \\ \hline \hline
			\multicolumn{9}{c}{$P_{e^-}$=-0.8} \\\hline\hline
			%$\sigma_{t}$ (fb)          &545.90&696.07&1238.31&1968.12&332.42&273.64&319.79&441.21  \\ \hline 
			$\sigma_{t}$ (fb)          &546&696&1238&1968&332&274&320&441  \\ \hline
			%$\sigma_{\bar{t}}$ (fb)           &289.20&295.26&1101.71&564.12&153.81&186.75&173.16&95.58  \\ \hline \hline
			$\sigma_{\bar{t}}$ (fb)           &289&295&1102&564&154&187&174&96  \\ \hline \hline
			\multicolumn{9}{c}{$P_{e^-}$=+0.8} \\\hline\hline
			%$\sigma_{t}$ (fb)          &514.78&414.44&1964.81&1238.32&195.53&229.78&299.85&233.90  \\ \hline
			$\sigma_{t}$ (fb)          &515&414&1965&1238&196&230&300&234  \\ \hline 
			%$\sigma_{\bar{t}}$ (fb)           &230.88&235.30&565.34&1102.11&128.59&109.86&72.07&116.52  \\ \hline \hline
			$\sigma_{\bar{t}}$ (fb)           &231&235&565&1102&129&110&72&117  \\ \hline \hline
		\end{tabular}
	}
	\caption{\small The partonic cross section of the signal process : $p e^- \to e^- tj, (t \to Wb, W \to \ell \nu_\ell)$ for different $Vtu$ FCNC couplings. The value of the anomalous couplings are set to unity with the cut off $\Lambda=m_t$, mass of top quark. Beam energies of $E_{e(p)}=60~(50000)$ GeV are considered.}
	\label{tab:CH_5_sig_cs} 
\end{table}

In all cases of $u$-quark coupling, the top quark production cross section is about two times larger than that of the top antiquark production. This is mainly due to the difference in the PDF of the anti quark in proton compared to that of the quark. On the other hand, the $c$-quark couplings lead to more or less the same cross section for both top quark and top antiquark. This is also similar to the case of top antiquark cross section in the case of $u$-quark couplings, indicating the role play by quark (or antiquark) PDF's in the cross section. The electron beam polarisation is another influencing factor at this collider. The polarisation of the electron beam influence the reactions differently depending on the type of anomalous couplings. We shall exploit this fact to distinguish the couplings. In this study we shall focus on the $tqg$ and $tq\gamma$ couplings, which spare better compared to the $tqZ$ couplings.

\begin{table}[tph]
	\centering
	\resizebox{0.99\linewidth}{!}
	%\small
	{
		\begin{tabular}{c|r|r||r|r||r|r||r|r} \hline
			Cooupling & $\kappa_{gct}^L$ & $\kappa_{gct}^R$ & $\kappa_{\gamma ct}^L$ & $\kappa_{\gamma ct}^R$ & $X_{zct}^L$  & $X_{zct}^R$ & $\kappa_{zct}^L$  & $\kappa_{zct}^R$  \\ \hline \hline
			\multicolumn{9}{c}{unpolarized} \\\hline\hline
			%$\sigma_{t}$ (fb)          &249.22&272.95&712.99&707.95&124.06&118.34&81.72&96.60  \\ \hline \hline
			$\sigma_{t}$ (fb)          &249&273&713&708&124&118&82&97  \\ \hline \hline
			%$\sigma_{\bar{t}}$ (fb) &244.54&248.96&709.93&707.11&118.58&123.67&96.71&81.79  \\ \hline \hline
			$\sigma_{\bar{t}}$ (fb) &245&249&710&707&119&124&97&81  \\ \hline \hline
			\multicolumn{9}{c}{$P_{e^-}$=-0.8} \\\hline\hline
			%$\sigma_{t}$ (fb)          &222.26&359.06&466.68&951.34&155.37&129.37&70.30&139.30  \\ \hline \hline
			$\sigma_{t}$ (fb)          &222&359&467&951&155&129&70&139  \\ \hline \hline
			%$\sigma_{\bar{t}}$ (fb)           &270.92&277.01&953.73&465.11&129.70&155.03&139.47&70.27  \\ \hline \hline
			$\sigma_{\bar{t}}$ (fb)           &271&277&954&465&130&155&139&70  \\ \hline \hline
			\multicolumn{9}{c}{$P_{e^-}$=+0.8} \\\hline\hline
			%$\sigma_{t}$ (fb)          &276.18&186.85&959.31&464.56&92.75&107.31&93.15&53.89  \\ \hline \hline
			$\sigma_{t}$ (fb)          &276&187&959&465&93&107&93&54  \\ \hline \hline
			%$\sigma_{\bar{t}}$ (fb)           &218.17&220.91&466.12&949.10&107.45&92.30&53.96&93.31  \\ \hline \hline
			$\sigma_{\bar{t}}$ (fb)           &218&221&466&949&107&92&54&93  \\ \hline \hline
		\end{tabular}
	}
	\caption{\small The partonic cross section of the signal process : $p e^- \to e^- t j, (t \to Wb, W \to \ell \nu_\ell)$ for different $Vtc$ FCNC couplings.  The value of the anomalous couplings are set to unity with the cut off $\Lambda=m_t$, mass of top quark. Beam
		energies of $E_{e(p)}=60~(50000)$ GeV are considered.}
	\label{tab:CH_5_sig_cs_cquark} 
\end{table}

Coming to the background, we present the cross sections of listed potential processes in \cref{tab:CH_5_bkg_cs}. Notice that choosing right-handed electron beam polarisation can reduce the backgrounds by a factor of 2 to 5, depending on the background process. 

\begin{table}[tph]
	\centering \small
	%\resizebox{0.95\textwidth}{!}{
	\begin{tabular}{c|r|r|r} \hline\hline
		&\multicolumn{3}{c}{$\sigma$ (fb)}\\\cline{2-4}
		Process & unpol & $P_e=-0.8$& $P_e=+0.8$ \\\hline\hline
		%$e^-jjl^+\nu_\ell$& 137.97 & 202.62 & 73.32 \\\hline
		$e^-jj\ell^+\nu_\ell$& 138 & 203 & 73 \\\hline
		%$e^-bjjl^+\nu_\ell$&5.68&8.00&3.36 \\\hline
		$e^-bjj\ell^+\nu_\ell$&6&8&3 \\\hline
		%$bjjj\nu_\ell$&43.84&78.90&8.77 \\\hline
		$bjjj\nu_\ell$&44&79&9 \\\hline
		%$bjj\nu_\ell$&144.32&259.77&28.86 \\\hline
		$bjj\nu_\ell$&144&260&29 \\\hline
		$e^-jj\ell^-\nu_\ell$&153&240&69\\\hline\hline
	\end{tabular}
	%	}
	\caption{\small The partonic cross section of the background processes (for both $t$ and $\bar t$ in the signal process) at different beam polarizations, with beam energies of $E_{e(p)}=60~(50000)$ GeV. Both $e$ and $\mu$ are included in $\ell$.}
	\label{tab:CH_5_bkg_cs} 
\end{table}

\section{Event Generation and Analysis} \label{eventgen}
\begin{figure}[tph]
	\centering
	\includegraphics[trim=0 0 0 0,clip,width=0.48\linewidth]{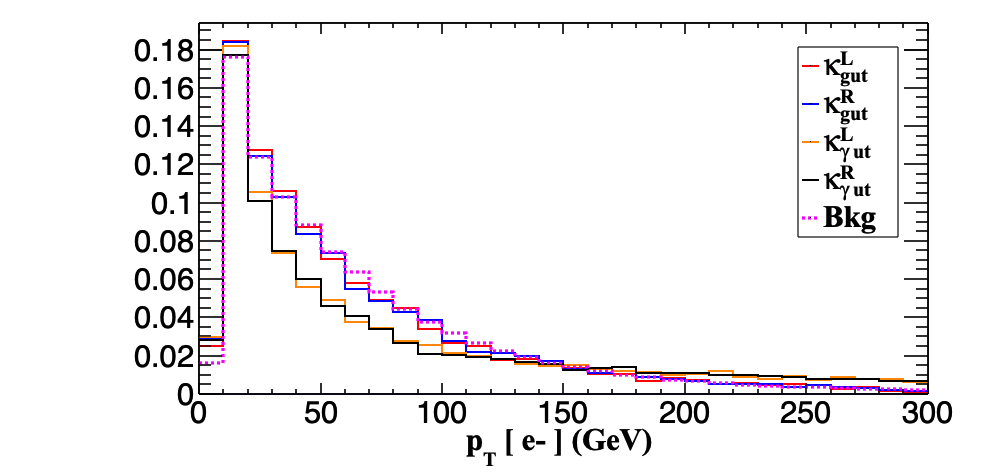}
	\includegraphics[trim=0 0 0 0,clip,width=0.48\linewidth]{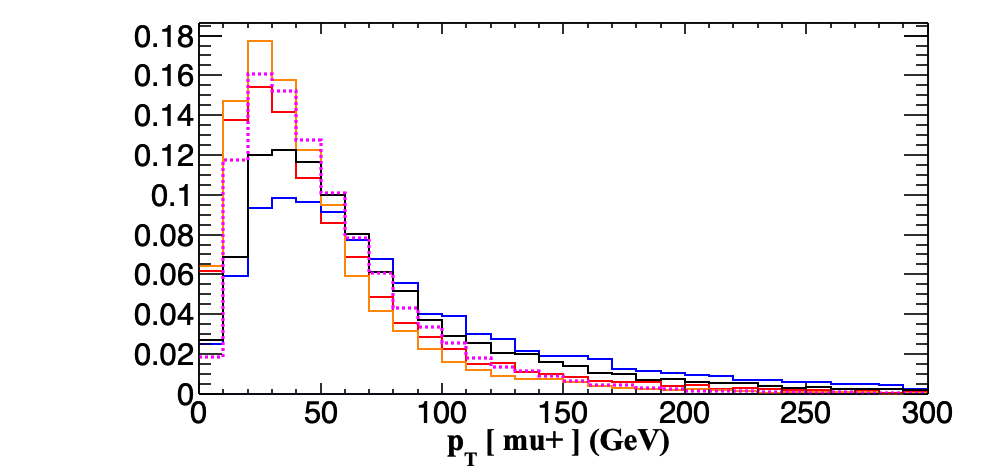}\\
	\includegraphics[trim=0 0 0 0,clip,width=0.48\linewidth]{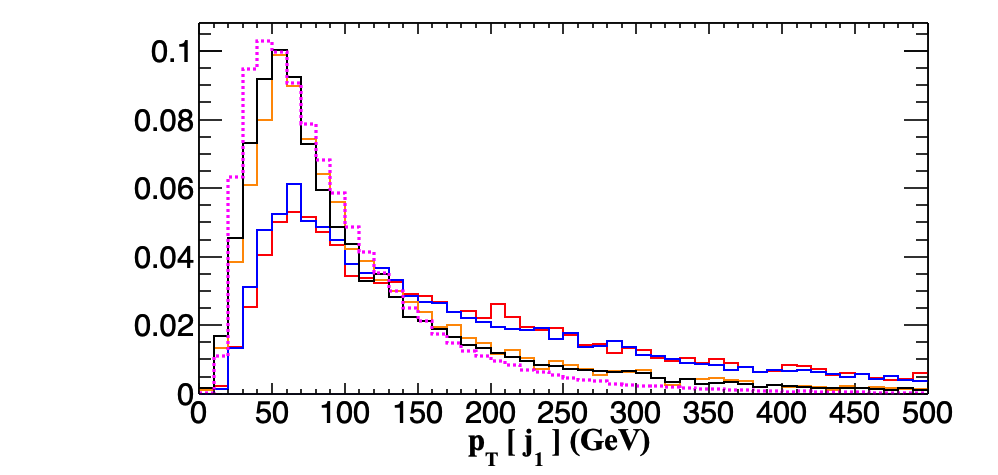}
	\includegraphics[trim=0 0 0 0,clip,width=0.48\linewidth]{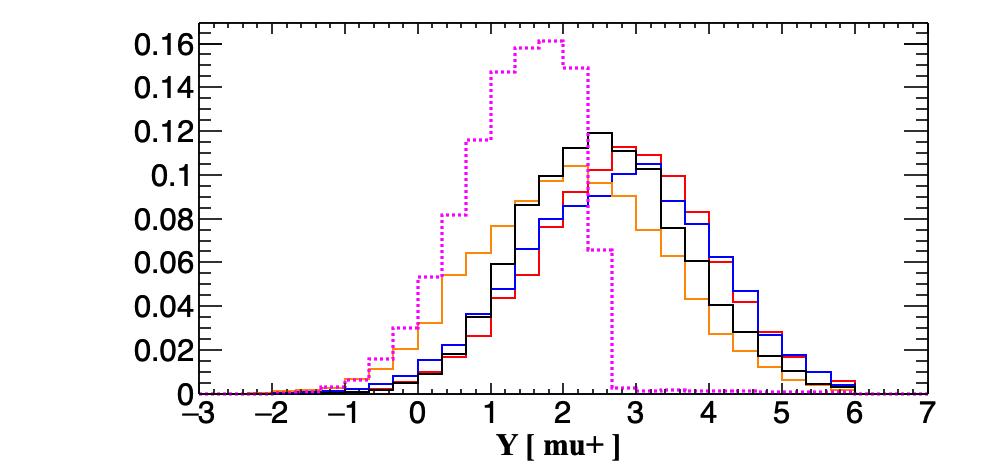}\\
	\includegraphics[trim=0 0 0 0,clip,width=0.48\linewidth]{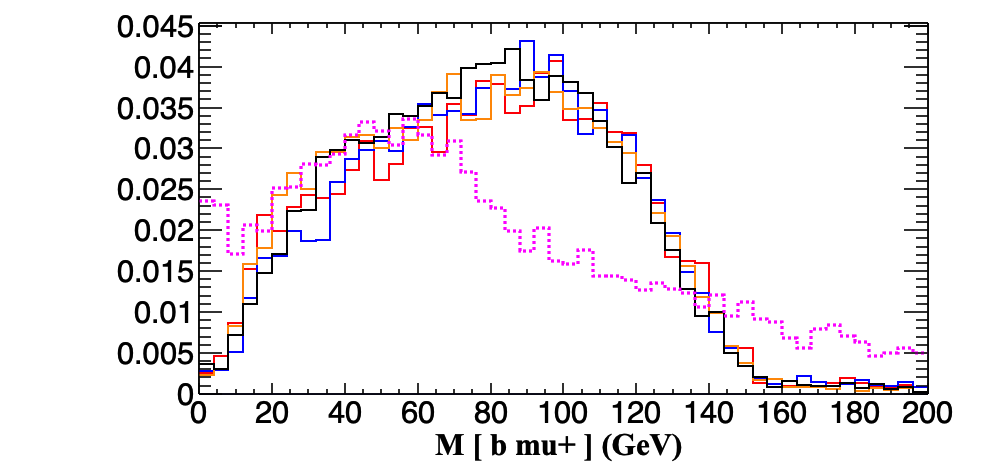}
	\includegraphics[trim=0 0 0 0,clip,width=0.48\linewidth]{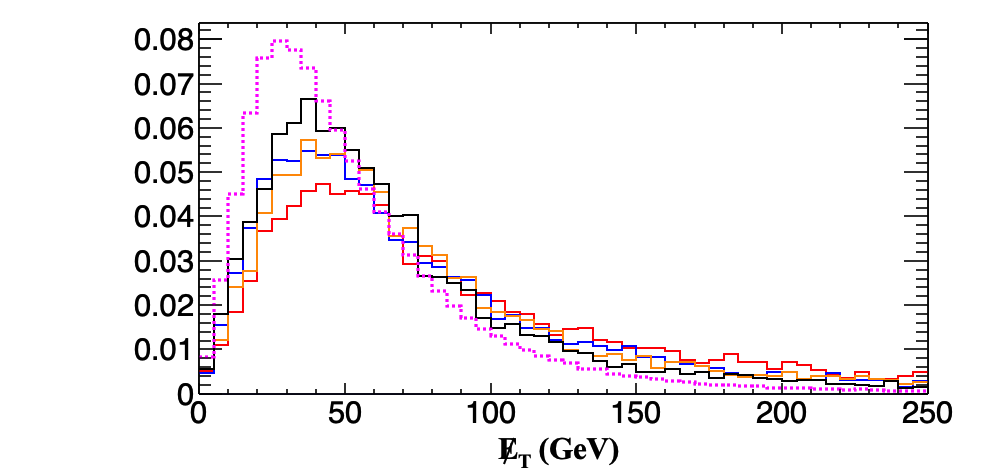}
	\caption{\small Some of the kinematic distributions for the case of $e^-p\to e^-tj\to e^-j\mu^+\nu$ after the application of basic cuts and selecting out events with $N_{e^-}=N_{\mu^+}=N_b=1$ and $N_j\ge 1$.The solid (dotted) line  represents the signals (background) events. Different case of signal are as indicated.}
	\label{fig:signalVsBkg_nopol}
\end{figure}

We use the FeynRules implementation of the effective Lagrangian considered in \cref{eq:Lag} in addition to the SM Lagrangian. For the event generation of both signal and background we use MADGRAPH5 \cite{Alwall:2014hca}, with a customised Pythia-PSG \cite{Sjostrand:2006za} performing the hadronization and showering. We used CTEQ6L1 PDF set with factorization and renormalization scales taken as $m_t$.  
Generation level event selection of  transverse momentum $p_T>10$ GeV and pseudo rapidity $|\eta|<5$ are imposed on all jets and leptons. All the events thus generated are passed through FastJet \cite{Cacciari:2011ma} to form the jets, where we used the anti-$k_T$ algorithm with cone size of $R=0.4$. The detector is emulated through Delphes \cite{deFavereau:2013fsa} with detector card tuned to take into account the asymmetric nature of the collider along with the very high energy proton beam resulting in highly boosted final state products. The events thus obtained after passing through Delphes are further analyzed using MadAnalysis5 \cite{Conte:2012fm} and ROOT \cite{Antcheva:2009zz}. The pre-selection of events is performed with the basic selection criteria
\begin{itemize}
	\centering
	\item[ ] $P_T^i>10$ GeV,~~~~~~~$|\eta_i|<5$,~~~~~~~~$\Delta R (i, j) \ge 0.4$,
\end{itemize}
where $i,~j\equiv$ electron, light-jet, or b-jet. Defined as usual $\Delta R=\sqrt{\Delta \eta^2 + \Delta \Phi^2}$, where $\eta$ is the pseudo-rapidity and $\phi$ is the azimuthal angle. 
In the case of signal, we have considered the SM decay of top quark which further gives either a $\mu^+$ or $e^+$, and the top antiquark into $\mu^-$ final state only. With the scattered electron always present, we have not considered the case of top antiquark decaying into electron, as the top quark reconstruction will be less efficient in this case.
The above basic selection of events is followed by the signal specific event selection demanding 
\begin{enumerate}
	%\centering
	\item one isolated $e^-$, %\\[-8.5mm]
	\item one isolated $\ell^+$ (either $e^+$ or $\mu^+$ for top quark, and $\mu^-$ for top antiquark production) %\\[-8.5mm]
	\item one $b$-jet,  %\\[-8.5mm]
	\item at least one light-jet. 
\end{enumerate}
This selection cuts down the signal by a factor of 2, whereas the background is reduced by a factor of 20. The cross sections before and after the selection are given in \cref{tab:cutflow_sig_unpol} for the case of unpolarised beam. Cases with the left- and right-beam polarisation are affected by this selection in a similar way. Only the  $ejj\ell\nu$ background is significant after this selection.  
For further event selection to enhance the signal over background we studied the kinematic distributions. In \cref{fig:signalVsBkg_nopol} some of the selected kinematic distributions are presented for the unpolarised electron beam, and considering the $\mu^+$ decay channel of the top quark. All other cases of $e^+$ and $\mu^-$ in the final state, and with left- or right-polarised electron beams have similar distributions. We have analysed those cases separately, and the final results in each case shall be presented towards the end. 

%The distributions for process having one $e^+$ also exhibits the same kinematic distributions. We have also examine the processes having one $\mu^+$ or $e^+$ in the case of $tgc$ or $t\gamma c$ anomalous couplings to have the similar shape of the distributions. This is expected because the process involving u- or c-quark FCNC couplings are differ by the cross sections only. 
Main distinguishing factor between the signal and the background is their prominent difference in the rapidity distribution of the decay lepton.
While the muon in the signal process come from the decay of the top quark through the $W$ boson, their source in the background are mostly from the vector boson fusion channels.  Coming from the top quark, muons in the signal are also boosted compared to the ones in the background.
The difference in the topology, and the asymmetric nature of the collider leads to different rapidity distributions. We shall exploit this to employ suitable selections to enhance the signal significance. Thus we select a rapidity region of $Y_\ell > 2.6$ bringing down the signal by another factor of 2- 4, depending on the nature of the couplings considered and the polarization of the beam used (see \cref{tab:cutflow_sig_unpol}).
%There is a slight dependence on the final state lepton (decay production of top quark) as well. 
The background, on the other hand, is considerably reduced with corresponding enhancement in the signal significance. The other two cases of  beam polarisation with $P_e=\pm 80\%$ show very similar behaviour with the selection.

\begin{table}[tph]
\centering
\resizebox{0.99\linewidth}{!}{
	\begin{tabular}{c|r|r|r|r||r|r|r|r||r} \hline
		Coupling&$\kappa_{gut}^L$&$\kappa_{gut}^R$&$\kappa_{\gamma ut}^L$&$\kappa_{\gamma ut}^R$&$\kappa_{gct}^L$&$\kappa_{gct}^R$&$\kappa_{\gamma ct}^L$&$\kappa_{\gamma ct}^R$&$Bkg$ \\\hline\hline
		\multicolumn{9}{c||}{ $e^-p\to e^-t j\to e^-bj\ell^+\nu$  (both $e^+$ and $\mu^+$ included)}&
		\\ \cline{1-10}
		%			Basic Cuts                         &530.6&	556.2&	1601.0&	1595.0&	249.3&	273.8&	710.0&	710.0&287.2 \\\hline
		%			$N_{e^-,\mu^+,b}=1,N_j \geqslant 1$&269.2&	291.4&	835.9  & 844.1	&128.4 &  142.9&  378.2&  386.8&4.1 \\\hline
		%			$Y_{\mu^+}>2.6$               &152.6&	154.6&	278.0  & 377.5	&55.4   &    42.9&    79.5&	 117.0&0.011 \\\hline\hline
		Basic Cuts                         &1058.0&	1113.2&	3198.0&	3190.0&	498.5&	546.2&	1420.2&	1420.0&485.0 \\\hline
		$N_{e^-}=N_{\ell^+}=N_b=1,~N_j \geqslant 1$&515&	550.0&	1631.7  & 1623.2	&245.9 &  271.5&  728.1&  737.0&8.5 \\\hline
		$Y_{\mu^+}>2.6$               &290.1&	287.6&	530.3 & 715.8&105.4  &    82.8&    155.9&	 219.0&0.023 \\\hline\hline
		
		\multicolumn{9}{c||}{$e^-p\to e^-\bar t j\to e^- j\mu^-\nu$}&
		\\ \cline{1-10}
		Basic Cuts                         &259.4	&266.1&	831.7&	829.1&	244.0&	248.8&	709.4&	708.9&485.0 \\\hline
		$N_{e^-}=N_{\mu^-}=N_b=1,~N_j \geqslant 1$&146.4	&150.1&	467.7&	480.1&	130.6&	128.8&	374.1&	376.6&4.8 \\\hline
		$Y_{\mu^-}>2.6$               &65.8	&53.6&	102.5&	151.0&	58.3&	47.2&	79.6&	118.2&0.009 \\\hline\hline
		
		%			\multicolumn{9}{c||}{$e^+$ Sample}&
		%			\\ \cline{1-10}
		%			Basic Cuts                         &527.4&	557.0&	1597.0&	1595.0&	249.3&	272.4&	710.2&	710.0&287.2 \\\hline
		%			$N_{e^-,e^+,b}=1,N_j \geqslant 1$&246.7&	258.6&	795.8&	779.1&	117.5&	128.6&	349.9&	350.2&4.4 \\\hline
		%			$Y_{e^+}>2.6$               &137.5&	133.0&	252.3&	338.3&	50.0&	39.9&	76.4&	102.0&0.011 \\\hline\hline
	\end{tabular}
}
\caption{Cross sections ($\sigma$ fb) with unpolarised electron beam at different selection levels.  The signal is considered with coupling values taken as unity, and the background is denoted as $Bkg$.}	\label{tab:cutflow_sig_unpol} 
\end{table}

In the case of top quark production, the fiducial cross sections of the signal events range from 300 to 700 fb, considering individual decay channels of the top quark and for unit couplings with the energy scale set to mass of the top quark ($\Lambda=m_t$).  For top anti-quark we consider only the $\mu^-$ channel, with the cross section after selection in the range of 50 to 150 fb. For more realistic coupling values of $\frac{\kappa^{L,R}}{\Lambda}=1 - 2~ \times 10^{-2} / m_t$, the cross section will be reduced by a factor corresponding to the square of the coupling. The best case scenario of $\ell^+$ final state is listed in \cref{tab:sensitivity}, taking the coupling to be $\frac{\kappa^2}{\Lambda^2}=\frac{2\times10^{-4}}{m^2_t}$, along with the signal sensitivity reachable at 2 ab$^{-1}$ luminosity.

\begin{table}[tph]
	\centering
	\resizebox{0.99\linewidth}{!}
	%\small 
	{
		\begin{tabular}{c|r|r|r|r||r|r|r|r} \hline
			$\kappa^2=\frac{2\times10^{-4}}{m_t^2}$&$\kappa_{gut}^L$&$\kappa_{gut}^R$&$\kappa_{\gamma ut}^L$&$\kappa_{\gamma ut}^R$&$\kappa_{gct}^L$&$\kappa_{gct}^R$&$\kappa_{\gamma ct}^L$&$\kappa_{\gamma ct}^R$ \\\hline\hline
			\multicolumn{9}{c}{unpolarized beam. Background events, $N_B=46$}\\\hline
			Signal, $N_S$ &116&115&212&286&42&33&62&88\\\hline
			Significance&9.1&	9.0&	13.2&	15.7&	4.5&	3.7&	6.0&	7.6\\\hline\hline
			\multicolumn{9}{c}{$P_e=-0.8$. Background events, $N_B=34$} \\\hline
			Signal, $N_S$&128&135&197&324&52&70&74&157\\\hline
			Significance&10.1&	10.4&	13.0&	17.1&	5.6&	6.9&	7.1&	11.3\\\hline\hline			
			\multicolumn{9}{c}{$P_e=+0.8$. Background events, $N_B=28$ }\\ \hline
			Signal, $N_S$   &109&125&195&328&56&42&116&95\\\hline
			Significance&9.3&	10.1&	13.0&	17.4&	6.1&	5.0&	9.6&	8.6\\\hline
		\end{tabular}
	}
	\caption{\small Signal significance with an assumed value of the coupling, $\frac{\kappa^2}{\Lambda^2}=\frac{2\times10^{-4}}{m^2_t}$ for the process $e^-p\to e^-tj\to e^-bj\ell^+\nu$ ($\ell^+$ includes both $e^+$ and $\mu^+$) . An integrated luminosity of 2 ab$^{-1}$ is considered.}   \label{tab:sensitivity} 
\end{table}

The significance is computed using the formula
\begin{equation}
{\rm Significance}~~=\frac{N_S}{\sqrt{N_S+N_B}}.  \label{eq:significance}
\end{equation}
\begin{figure}[tph]
	\centering
	\includegraphics[width=0.60\linewidth]{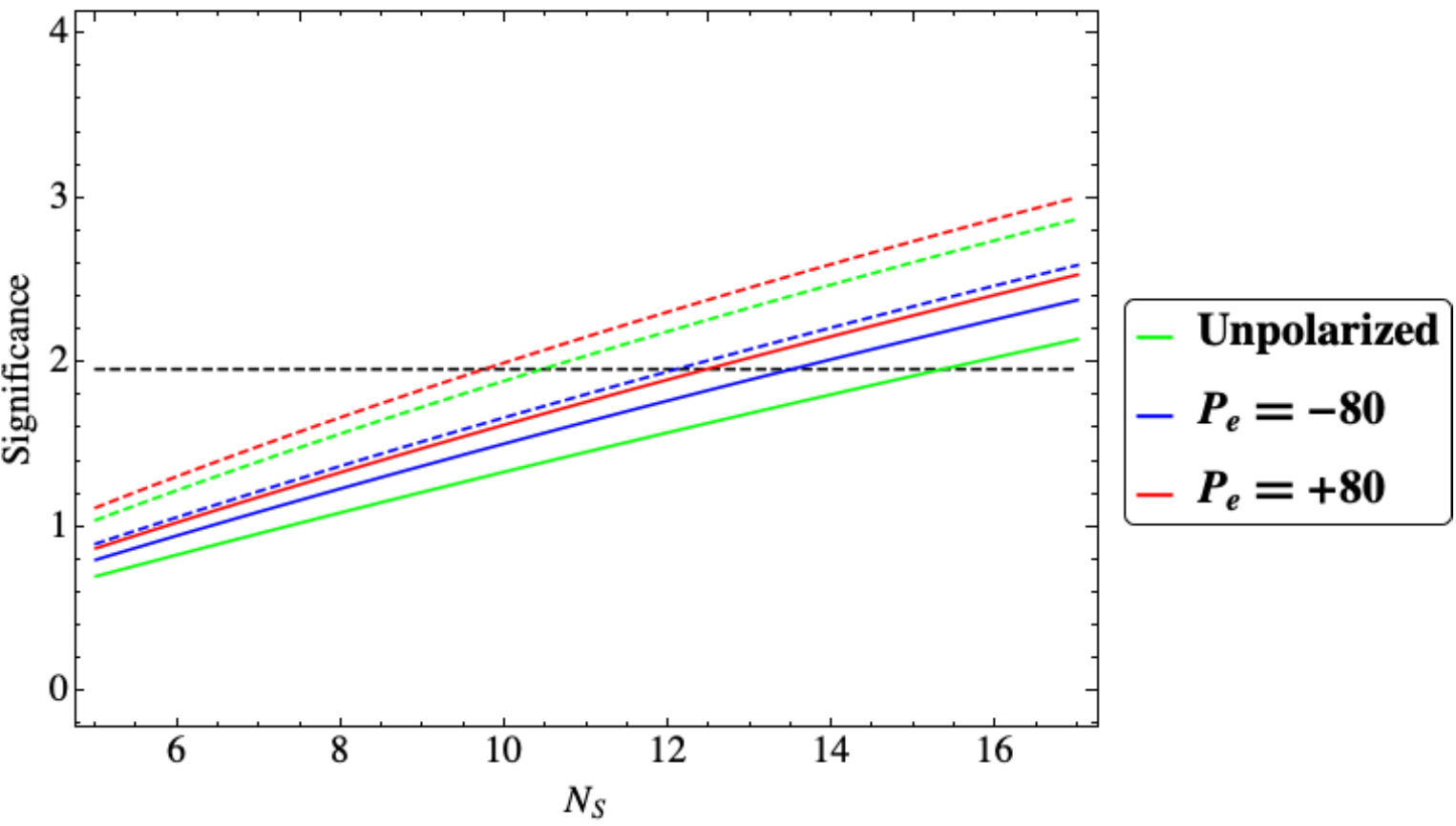}
	\caption{Significance against the number of signal events required for specific number of background events as per Eq.~\ref{eq:significance}. Number of background events corresponding to the cases of $e^-p\to e^-tj\to e^-bj\ell^+\nu$ including both $e^+$ and $\mu^+$ in the final state (solid lines), and that corresponding to the case of $e^-p\to e^-\bar tj\to e^-j\mu^-\nu$ (dotted lines) considered for the three beam polarisations.}
	\label{plot:significance}
\end{figure}
The right polarised beam shows a marginally better significance as compare to other two samples. To obtain the reach of the coupling at a given confidence level we use the plot for significance versus the number of signal events \cref{plot:significance}. Both the cases of $t\to b\ell^+\nu_\ell$ (solid curve) adding $\mu^+$ and $e^+$ events, and $\bar t \to b\mu^-\bar \nu_\mu$ (dotted curve) are presented for three different electron beam polarisations considered, along with the black dashed line  indicating the  $95\%$ C.L. value.
\begin{table}[tph]
	\centering
	\resizebox{0.99\linewidth}{!}
	{
		\begin{tabular}{c|c||c|c|c|c|c|c}
			\hline
			Beam&Associated&Background&Required&\multicolumn{4}{c}{Reach of couplings for $95\%$ CL. (TeV$^{-1}$)}\\ \cline{5-8}
			
			polarization&quark&events, $N_B$&signal, $N_S(2\sigma)$
			&$\kappa_{gqt}^L/\Lambda$&$\kappa_{gqt}^R/\Lambda$&$\kappa_{\gamma qt}^L/\Lambda$&$\kappa_{\gamma qt}^R/\Lambda$ \\\hline \hline
			
			\multirow{2}{*}{Unpolarized}&$up$&\multirow{2}{*}{46}&\multirow{2}{*}{16}
			&0.029&	0.029&	0.021&	0.019  \\ \cline{2-2}\cline{5-8}
			&$charm$&&&0.048&	0.054&	0.040&	0.033\\ \hline\hline
			
			\multirow{2}{*}{$(-80\%)$}&$up$&\multirow{2}{*}{34}&\multirow{2}{*}{14}
			&0.028&	0.027&	0.022&	0.017	\\ \cline{2-2}\cline{5-8}
			&$charm$&&&0.043&	0.037&	0.036&	0.025\\ \hline\hline
			
			\multirow{2}{*}{$(+80\%)$}&$up$&\multirow{2}{*}{28}&\multirow{2}{*}{13}
			&0.030&	0.030&	0.022&	0.019\\ \cline{2-2}\cline{5-8}
			&$charm$&&&0.050&	0.056&	0.041&	0.035 \\ \hline
		\end{tabular}
	}
	\caption{\small Number of signal events required for $2\sigma$ significance (refer to \cref{plot:significance}), and the 95\% C.L. reach on the corresponding couplings for different beam polarizations in $e^- p \to e^-tj\to e^-bj\ell^+\nu$ (with $\ell^+$ including both $e^+$ and $\mu^+$) at an integrated  luminosity of 2 ab$^{-1}$.}
	\label{table:reach}
\end{table}
The number of events required to have 95\% C.L., for the three different cases of initial electron beam polarisations are given in \cref{table:reach}, along with the corresponding reach on the couplings, setting the cut-off $\Lambda=1$ TeV.  The branching ratios of the corresponding rare channels in the best case scenario of $-80\%$ beam polarisation is comparable to, and in some cases marginally better than that achievable at the HL-LHC (notice that we consider smaller integrated luminosity than what is considered in the case of HL-LHC), which is given in \cref{tab:brreach}. The cases of unpolarised beam and right-polarised beam have similar reach, with marginal differences. 
\begin{table}[tph]
	\centering
	%	\resizebox{0.95\linewidth}{!}{
	\small
	\begin{tabular}{c|c|c|c|c}
		\hline\hline
		Coupling&\multicolumn{4}{c}{{\rm BR} $\times 10^5$}\\ \cline{2-5}
		&$t\to ug$&$t\to cg$&$t\to u\gamma$&$t\to c\gamma$\\ \hline\hline
		$\kappa_{Vqt}^L/\Lambda$&1.8&4.3&0.74&1.9\\
		$\kappa_{Vqt}^R/\Lambda$&1.7&3.1&0.44&0.95\\ \hline\hline
	\end{tabular}
	\caption{\small 95\% C.L. reach on the branching fractions of the rare decay, corresponding to the reach on the relevant coupling (as in \cref{table:reach}) in the best case of $-80\%$ beam polarisation with 2 ab$^{-1}$ integrated luminosity.} 
	\label{tab:brreach}
\end{table}
Judging from the numbers in \cref{tab:cutflow_sig_unpol} the case of top antiquark is very similar to that of the $c$ quark couplings in top production case. Thus, for the $u$ quark coupling, top quark case has an edge over the top antiquark case, whereas for the $c$ quark couplings, both give similar results. Besides considering separately, one may perform a combined analyses of all the cases, which may increase the sensitivity significantly. However, we do not attempt this in the present discussion. 

\section{Top polarization Asymmetries} \label{ch_5:spin_matrix}
In this section we discuss the formalism that could be employed to extract the polarization information of the top quark through suitably constructed observables. For details of the formalism one may consult Ref.~\cite{Behera:2018ryv} and references therein. % Ref.~\cite{Godbole:2006tq, Boudjema:2009fz,Rindani:2011pk,}. 
%As explained in the previous section, 
The motivation for the spin analysis of top quark comes from the fact that the angular distributions of top quark decay products give access to the Lorentz structure of the production vertex through the information of top quark polarisation. To extract the information we proceed as follows.

In the Narrow Width Approximation (NWA), the invariant amplitude square of the full process ($eq\rightarrow etj \rightarrow ejb\ell\nu$) can be written as a product of the production and decay density matrices in the helicity basis of the top quark as

\begin{equation}\label{eq:3}
{|{\cal M}|^2}
=
\frac{\pi\delta(p_t^2 -m_t^2)}{\Gamma_t m_t}
\sum_{\lambda,\lambda'}\rho(\lambda,\lambda')\Gamma(\lambda,\lambda'),
\end{equation} 
where $p_t$ is the momentum and $\Gamma_t$ is the total width of the top quark, with the summation considered over the helicity indices of the top quark. The production and decay density matrices are given in terms of the corresponding amplitudes as $\rho(\lambda,\lambda') = {\cal M}_P(\lambda)\,{\cal M}_P^*(\lambda')$ and $\Gamma(\lambda,\lambda') = {\cal M}_\Gamma(\lambda)\,{\cal M}_\Gamma^*(\lambda')$, respectively. The top quark on-shell condition in the NWA allows one to write the differential cross section of  the complete process as ~\cite{Boudjema:2009fz}

\begin{align}
\frac{1}{\sigma_t}\frac{d\sigma_t}{d\Omega_t} = \frac{1}{4\pi}~\sum_{\lambda,\lambda'}\sigma(\lambda,\lambda')\Gamma(\lambda,\lambda')
% =\frac{1}{\sigma_{\rm prod}}\rho_T(\lambda,\lambda')
\end{align} 
where we define the normalised production density matrix of the top quark,
\begin{align}
\sigma(\lambda,\lambda')
=
\frac{1}{\sigma_{\rm prod}}\int \rho(\lambda,\lambda') ~d\Omega_t,
% =\frac{1}{\sigma_{\rm prod}}\rho_T(\lambda,\lambda')
\end{align} 
with $d\Omega_t$ representing the differential solid angle of top quark produced and $\sigma_{\rm prod}$ denoting the total production cross section. For convenience, we define polarisation vector ${\bf P}=(P_x,P_y,P_z)$ so that

\begin{align}
\begin{aligned}
\sigma(+,+)&=\frac{1}{2}(1+P_z),
&\sigma(+,-)&=\frac{1}{2}(P_x+iP_y),\\
\sigma(-,-)&=\frac{1}{2}(1-P_z),
&\sigma(-,+)&=\frac{1}{2}(P_x-iP_y).
\end{aligned}
\label{eq:sig_Py}
\end{align}
The normalized decay density matrix elements for the process $t\to W^+ b\to b \ell^+ \nu_\ell$ may be written in terms of the polar ($\theta_\ell$) and azimuthal ($\phi_\ell$) angles of the secondary lepton in the top rest frame as~\cite{Boudjema:2009fz},
\begin{align}
\begin{aligned}
\Gamma(+,+)&=\frac{1}{2} (1+ \cos \theta_\ell),
&\Gamma(+,-)&=\frac{1}{2} \sin \theta_\ell e^{i\phi_\ell}, \\ 
\Gamma(-,-)&=\frac{1}{2} (1- \cos \theta_\ell) ,
&\Gamma(-,+)&=\frac{1}{2} \sin \theta_\ell e^{-i\phi_\ell}.
\end{aligned}
\label{decay_matrix}
\end{align}
Here the polar angle is measured with respect to the top quark boost direction, and the top production plane is taken as the $x$-$z$ plane. These choices of reference do not cost us generality of the analysis as shown in Ref.~\cite{Godbole:2006tq}. The differential cross section for the complete process in terms of the top quark polarisation vector and the polar and azimuthal angle of the secondary lepton in the rest frame of the top quark, can now be written as
\begin{align}
\frac{1}{\sigma_{t}}\frac{d\sigma_t}{d\Omega_\ell}
=
\frac{1}{4\pi} \Big(1+\,P_z~ \cos\theta_\ell + P_x ~\sin\theta_\ell ~\cos\phi_\ell
\nonumber\\
+\, P_y ~\sin\theta_\ell ~\sin\phi_\ell \Big),
\end{align}
where $\Omega_\ell$ is the solid angle of the secondary lepton. This enables one to define angular asymmetries of the secondary leptons, and connect those directly to the top quark polarisation. The following three asymmetries, two defined in terms of the azimuthal angle, and one in terms of the polar angle of the decay lepton, are used in the subsequent study.
\begin{gather}
\begin{aligned}
A_x \equiv\,&
\frac{1}{\sigma_{\rm tot}}\bigg[
\int_{-\frac{\pi}{2}}^{\frac{\pi}{2}} \!\! d\phi_\ell \frac{d\sigma}{d\phi_\ell}
- \int_{\frac{\pi}{2}}^{\frac{3\pi}{2}} \!\! d\phi_\ell \frac{d\sigma}{d\phi_\ell}
\bigg] =\frac{1}{2} P_x , \\
A_y \equiv\,&
\frac{1}{\sigma_{\rm tot}}\bigg[
\int_{0}^{\pi} \!\! d\phi_\ell \frac{d\sigma}{d\phi_\ell}
- \int_{\pi}^{2\pi} \!\! d\phi_\ell \frac{d\sigma}{d\phi_\ell}
\bigg] =\frac{1}{2} P_y , \\
A_z \equiv\,&
\frac{1}{\sigma_{\rm tot}}\bigg[
\int_{0}^{1} \!\! d\c_{\theta_\ell} \frac{d\sigma}{d\c_{\theta_\ell}}
- \int_{-1}^{0} \!\! d\c_{\theta_\ell} \frac{d\sigma}{d\c_{\theta_\ell}}
\bigg]=\frac{1}{2} P_z.
\end{aligned}
\label{eq_asym}
\end{gather}
With the couplings considered to be real, the asymmetry, $A_y$ is identically zero owing to Eq.~\ref{eq:sig_Py}.
Note that the angles in the above asymmetries are defined in the rest frame of the top quark, and thus require full reconstruction of the top quark momentum. In the present case this leads to the following relations between the components of the missing momentum (neutrino in this case) denoted by $p_{x\nu},~p_{y\nu},~p_{z\nu}$, and those of the visible final particles.
\begin{eqnarray}
p_{x\nu} &=& - \sum_{k=e,j,\ell,b} p_{x k},
\qquad p_{y\nu} = - \sum_{k=e,j,\ell,b} p_{y k}, \nonumber\\
(p_{z\nu})_\pm
&=&
\frac{1}{p_{T\ell}^2}
\Big[ \beta p_{z\ell} \mp E_\ell \sqrt{\beta^2 - p_{T\nu}^2 p_{T\ell}^2} \Big],
\end{eqnarray}
where $\beta = \frac{m_W^2}{2} + p_{x\ell} p_{x\nu} + p_{y\ell} p_{y\nu}$ and $p_{Ti}^2=p_{xi}^2+p_{yi}^2$. Out of the above two solutions for $p_{z\nu}$, the one for which $|\sum_r p_r^2 - m_t^2|$ is minimum, where $p_r$ is the four momentum of the corresponding particle, with $r=\ell,b,\nu$, will be considered as the correct choice for the $z$-component. The missing momentum thus obtained is used to reconstruct the top quark momentum.

\begin{figure}[tph]
	\centering
	\includegraphics[trim=0 0 0 0,clip,width=0.48\linewidth]{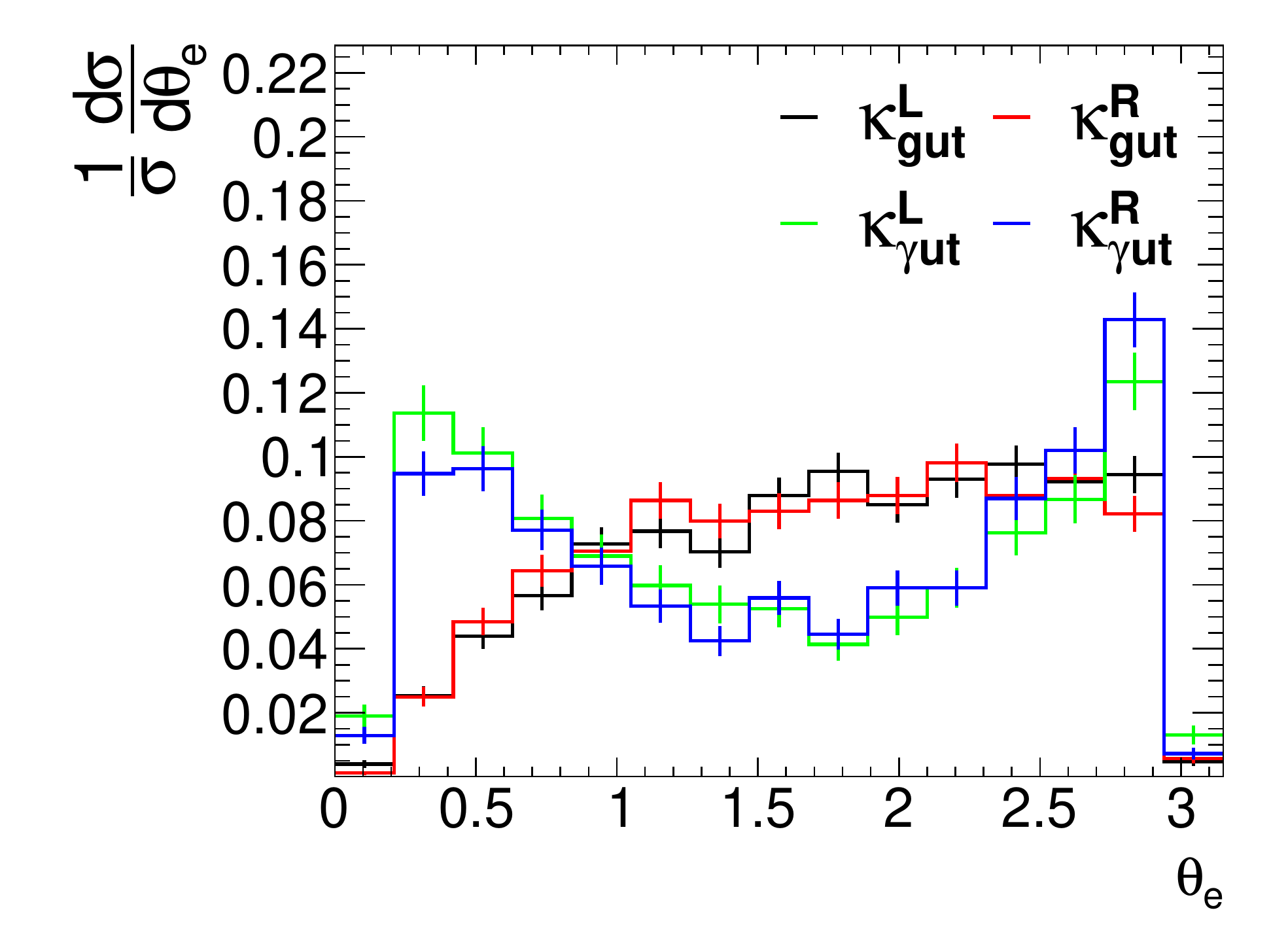}
	\includegraphics[trim=0 0 0 0,clip,width=0.48\linewidth]{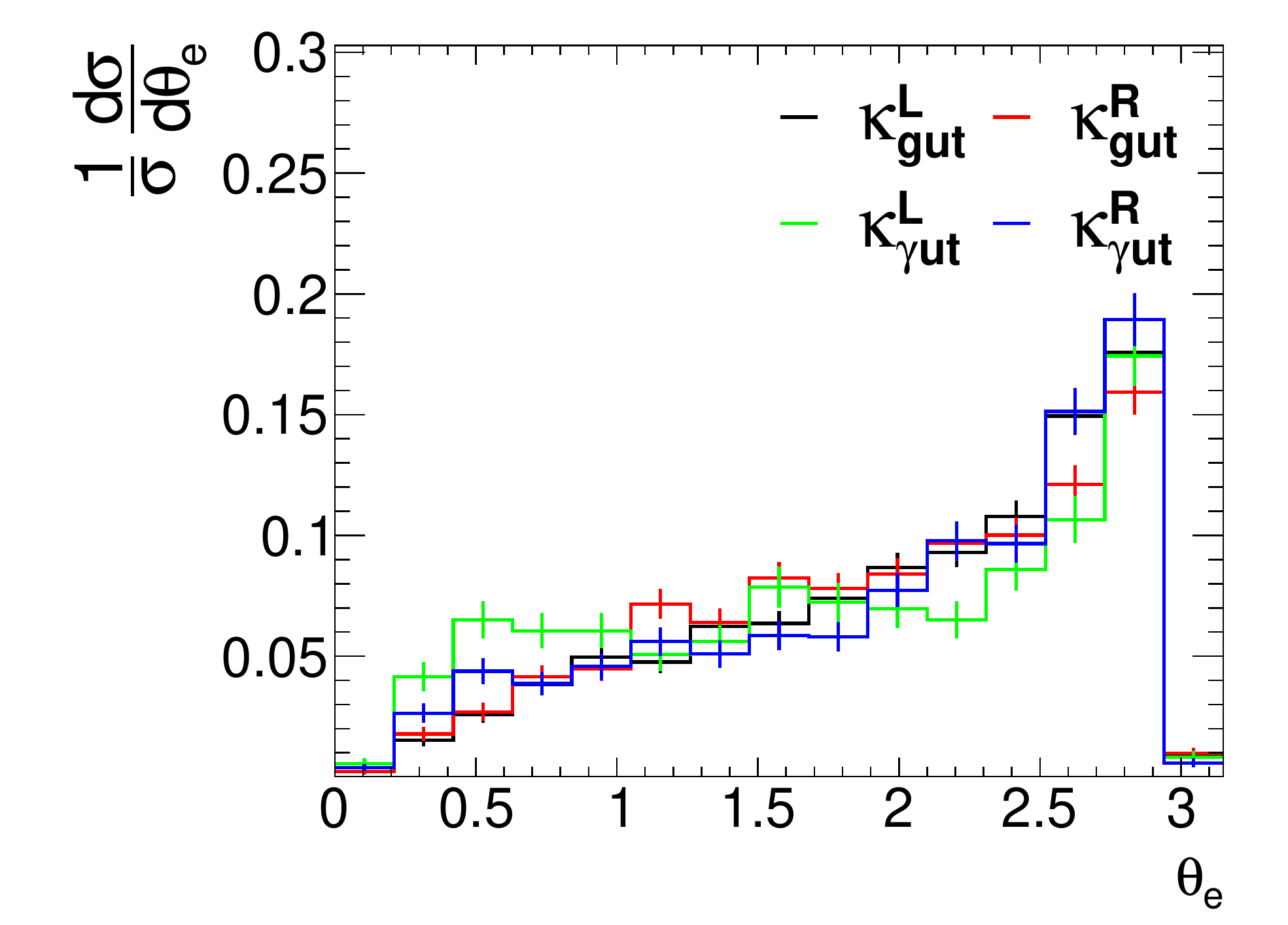}
	\caption{\small Angular distribution of the scattered electron in the case of unpolarised beam normalised to unity. The left figure corresponds to the top quark production, while the right one corresponds to the top antiquark production.} \label{fig:signal_asym}
\end{figure}

In addition to the asymmetries related to secondary lepton, we may exploit the advantage of producing the top quark in association with the electron in distinguishing the nature of the coupling.
In \cref{fig:signal_asym} we plot the angular distributions of the scattered electron in the lab frame of the collider with $z$-axis along the proton beam direction. As seen, the polar angle of the electron is sensitive to whether gluon coupling ($tgq$)  or photon coupling ($t\gamma q$) is considered. This is in a way intuitively clear, as the scattered electron interact with the quarks through photon or $Z$, and thus decoupled from the effects in the gluon interactions. 
%Since both photon and gluon interactions are non-chiral, the distributions are insensitive to the chirality of the coupling. 
On the contrary it seems to be insensitive to the case of anti top quark production,  indicating that the forward scattering of the electrons are dominated by the quark initiated processes. 
We define the forward-backward asymmetry,
\begin{align}
\centering
A_{e^-}^{FB} &=
\frac{\sigma(\cos{\theta_{e^-}} > 0)-\sigma(\cos{\theta_{e^-}} < 0)}{\sigma(\cos{\theta_{e^-}} > 0)+\sigma(\cos{\theta_{e^-}} < 0)},
%	\\
%	A_j^{FB} &=
%	\frac{\sigma(\cos_{\theta_j} > 0)-\sigma(\cos_{\theta_j} < 0)}{\sigma(\cos_{\theta_j} > 0)+\sigma(\cos_{\theta_j} < 0)},
\end{align}
where we denote the polar angle of the electron by $\theta_{e^-}$.
Numerical values of all the three asymmetries (i.e., $A_x$, $A_z$ and $A_{e^-}^{FB}$) for different couplings used are presented in \cref{ch_5_axayaz_Det_cut} in the case of top quark production with $\ell^+$ in the final state, and in \cref{ch_5_axayaz_Det_cut_1} in the case of top antiquark production with $\mu^-$ in the final state.

\begin{table}[tph]
	\centering
	\resizebox{0.99\linewidth}{!}
	%\small
	{
		\begin{tabular}{c|r|r|r||r|r|r||r|r|r} \hline
			&\multicolumn{3}{c||}{No pol.}&\multicolumn{3}{c||}{$P_{e^-} = - 80\%$}&\multicolumn{3}{|c}{$P_{e^-} = + 80\%$}
			\\ \cline{2-10}
			Coupling  & $A_x$ & $A_z$ & $A_{e^-}^{FB}$& $A_x$ & $A_z$ & $A_{e^-}^{FB}$ & $A_x$ & $A_z$ & $A_{e^-}^{FB}$  \\ \hline \hline
			
			$\kappa_{g ut}^L$             &-0.20&--&-0.22&-0.27&--&-0.28&-0.21&--&-0.20  \\\hline 
			$\kappa_{g ut}^R$            & 0.14& 0.40&-0.15&0.16&0.42&-0.23&0.11&0.41&-0.16   \\\hline 
			$\kappa_{\gamma ut}^L$  &-0.33&-0.12& --&-0.31&-0.11&--&-0.30&-0.16&--   \\\hline 
			$\kappa_{\gamma ut}^R$ &--& 0.46&--&--&0.47&--&--&0.48&--   \\\hline\hline
			
		\end{tabular}
	}
	\caption{\small Top polarisation asymmetries and forward-backward asymmetry of the scattered electron with three different beam polarisations, considering presence of one coupling at a time in the process $e^-p\to e^-\bar u t$ with $t\to b\ell^+ \nu$.}
	\label{ch_5_axayaz_Det_cut}
\end{table}

\begin{table}[tph]
	\centering
	\resizebox{0.99\linewidth}{!}
	%\small 
	{
		\begin{tabular}{c|r|r|r||r|r|r||r|r|r} \hline
			&\multicolumn{3}{c||}{No pol.}&\multicolumn{3}{c||}{$P_{e^-} = - 80\%$}&\multicolumn{3}{|c}{$P_{e^-} = + 80\%$}
			\\ \cline{2-10}
			Coupling  & $A_x$ & $A_z$ & $A_{e^-}^{FB}$& $A_x$ & $A_z$ & $A_{e^-}^{FB}$ & $A_x$ & $A_z$ & $A_{e^-}^{FB}$  \\ \hline \hline
			
			$\kappa_{g ut}^L$            &-0.35&0.26&-0.48&-0.32&0.28&-0.48&-0.34&0.19&-0.46   \\\hline 
			$\kappa_{g ut}^R$             & --&0.33&-0.39&--&0.41&-0.47&--&0.32&-0.35  \\\hline 
			$\kappa_{\gamma ut}^L$   &-0.44&--&-0.25&-0.43&--&-0.28&-0.45&--&-0.28  \\\hline 
			$\kappa_{\gamma ut}^R$  &-0.11&0.54&-0.44&-0.14&0.56&-0.37&--&0.53&-0.39   \\\hline\hline
		\end{tabular}
	}
	\caption{\small Top polarisation asymmetries and forward-backward asymmetry of the scattered electron with three different beam polarisations, considering presence of one coupling at a time in the process $e^-p\to e^- u \bar t$ with $\bar t\to \bar b\mu^- \bar \nu$.}	\label{ch_5_axayaz_Det_cut_1}
\end{table}

\subsection{Disentangling different types of couplings}

We shall turn to the possibility to identify the type of coupling present by making use of a combination of observables discussed above. To start with we may notice that  a comparison of event rates with different beam polarizations could  be useful here. To illustrate this, we shall consider the fiducial cross sections after the event selection quoted in \cref{tab:sensitivity} for top quark production and its leptonic decay.  We shall first consider the $u$ quark coupling, where going from unpolarised beam to $-80\%$ polarised beam, the cross section is increased by 10 - 18\% for the case of gluon couplings and right-handed photon coupling. In the case of left-handed photon coupling, it has decreased by 10\% in number of events.  Going from unpolarised to $+80\%$ polarised beams, only the right-handed gluon and photon couplings present an appreciable change by an increment of about 10\% and 15\% respectively, while all other cases registering a decrease of 10\% from the unpolarized samples.  On the other hand, switching the polarisation from $-80\%$ to $+80\%$ leaves all cases within the 10\%, except the left-handed gluon coupling, which shows a decrease of about 15\%. Coming to the case of $c$ quark coupling, the changes are more dramatic, indicating that the quark initiated processes are more sensitive to the electron beam polarisation. In this case, going from unpolarised beam to $-80\%$ beam polarisation  shows an increase in all cases, with about 20\% in case of left-handed gluon or photon couplings and around 80\% and 112\% in the case of right-handed gluon and photon couplings, respectively. Moving from unpolarised to $+80\%$ polarised beam, on the other hand, while showing increase in all cases, the left-handed photon coupling poses a large difference close to 90\%, while the right-handed photon case registering only small change. Switching the polarisation from $-80\%$ to $+80\%$ a considerable downward change in the case of right-handed couplings in both the cases (about 40\%), while the left-handed case is undemocratic for the gluon couplings whereas, the photon case showing a large increase of about 57\%. We summarise these qualitative features in the cartoon in \cref{tab:distinguish_pol}.

\begin{table}[tph]
	\centering
	\resizebox{0.99\linewidth}{!}{
		%\small{
		\begin{tabular}{c|c|c|c}	
			\multicolumn{4}{c}{\scriptsize ($\cdots~:~< 10\%$ change), ($\downarrow$ or $\uparrow~:~\sim10\% - 50\%$), ($\Downarrow$ or $\Uparrow~:~>50\%$)}\\
			\hline\hline
			&{\rm unpol $\rightarrow -80$\%}&{\rm unpol $ \rightarrow +80$\%}&{\rm $-80\% \rightarrow +80$\%}\\ \hline
			\hline
			\multicolumn{4}{c}{$u$ quark couplings}\\
			\hline
			$\kappa^L_{gut}$&$\uparrow$&$ \cdots$&$\downarrow$\\\hline
			$\kappa^R_{gut}$&$\uparrow$&$ \cdots$&$ \cdots$\\ \hline
			$\kappa^L_{\gamma ut}$& $\cdots$&$ \cdots$& $\cdots$\\\hline
			$\kappa^R_{\gamma ut}$&$\uparrow$&$\uparrow$&  $\cdots$\\\hline
			\multicolumn{4}{c}{$c$ quark couplings}\\
			\hline
			$\kappa^L_{gct}$&$\uparrow$&$ \uparrow$&$\cdots$\\\hline
			$\kappa^R_{gct}$&$\Uparrow$&$ \uparrow$&$ \downarrow$\\ \hline
			$\kappa^L_{\gamma ct}$& $\uparrow$&$ \Uparrow$& $\Uparrow$\\\hline
			$\kappa^R_{\gamma ct}$&$\Uparrow$&$\cdots$&  $\downarrow$\\\hline\hline
	\end{tabular}}
	\caption{Demonstrating the power of beam polarization to disentangle the effects of different types of couplings. The up- or down-arrows 
		indicate an increase or decrease in the number of events.  For the number of events at 2 ab$^{-1}$ luminosity, see \cref{tab:sensitivity}.} 
	\label{tab:distinguish_pol}
\end{table}

In the following we discuss how the qualitative features of the top polarisation asymmetries ($A_x$, $A_z$) and electron forward-backward asymmetry ($A^{FB}_{e^-}$) are useful in discriminating the type of coupling. In \cref{v_t_separation_t_tbar} we capture these features in the case of top/antitop quark production.
 The uncertainty in the determination of asymmetries are assumed to be within 10\% level, and thus ignored those which measure less than 10\% (indicated as ``$\cdots$'' in \cref{v_t_separation_t_tbar}). In the case of top quark production, the forward-backward asymmetry of the scattered electron in the case of photon couplings is within the uncertainty. The $A_z$ asymmetry is positive and large (close to 50\%) in the case of right-handed couplings. Thus, presence of negative $A^{FB}_{e^-}$ along with positive $A_z$ will indicate the presence of $\kappa_{gut}^R$. If $A_z$ is absent, on the other hand, the coupling responsible is $\kappa_{gut}^L$.  If $A^{FB}_{e^-}$ is absent and $A_z$ is large and positive it indicates $\kappa_{\gamma ut}^R$, whereas negative $A_z$ indicates the presence of $\kappa_{\gamma ut}^L$.  Presence or absence of $A_x$ along with its sign can be used as additional marker. 
These qualitative distinctions comparing the asymmetries are independent of the beam polarization. 
\begin{table}[tph]
	\centering
	%\resizebox{0.99\linewidth}{!}
	\small
	{
		\begin{tabular}{c|c|c|c}
			\multicolumn{4}{c}{From top quark}\\ \hline
			$A_x$&$A_z$& $A_e^{FB}$ & Coupling \\ \hline\hline
			$-$ve &$\cdots$&$-$ve  & $\kappa_{g ut}^L$ \\ \hline
			$+$ve&$+$ve &$-$ve & $\kappa_{g ut}^R$ \\ \hline
			$-$ve &$-$ve &$\cdots$ &$\kappa_{\gamma ut}^L$ \\ \hline
			$\cdots$&$+$ve &$\cdots$&$\kappa_{\gamma ut}^R$ \\ \hline\hline
		\end{tabular}
		\hskip 5mm
		\begin{tabular}{c|c|c|c}
			\multicolumn{4}{c}{From top antiquark}\\ 	 \hline
			$A_x$&$A_z$& $A_e^{FB}$ & Coupling \\ \hline\hline
			$-$ve &$+$ve &$-$ve  & $\kappa_{g ut}^L$ \\ \hline
			$\cdots$&$+$ve &$-$ve & $\kappa_{g ut}^R$ \\ \hline
			$-$ve &$\cdots$ &$-$ve &$\kappa_{\gamma ut}^L$ \\ \hline
			$-$ve &$+$ve &$-$ve &$\kappa_{\gamma ut}^R$ \\ \hline\hline
		\end{tabular}
	}
	\caption{ Qualitative properties of the top polarization asymmetries and the forward-backward asymmetry of the scattered electron, allowing one to distinguish the effects.}	\label{v_t_separation_t_tbar}
\end{table}
In the case of top antiquark, the scattered electron does not help. However, between $A_x$ and $A_z$ partial discrimination is achievable.  In all cases $A_x$ is negative and $A_z$ is positive. However,  absence of $A_z$ and large negative $A_x$ indicates the presence of $\kappa_{\gamma ut}^L$ in an unambiguous manner. Switching this, that is, when large $A_z$ is present in the absence of $A_x$, indicates $\kappa_{g ut}^R$. Qualitatively, $\kappa_{g ut}^L$ and $\kappa_{\gamma ut}^R$ shows similar behaviour, although there are some quantitative distinctions.

\section{Summary and Conclusion} \label{ch_5conclusion}
The possibility of measuring FCNCs connected to top quark will clearly indicate dynamics beyond that of the SM. While the SM predicts very tiny effects arising through higher order quantum corrections, many popular extensions of the SM indicate possibility of much larger value for these couplings, close to what could be explored at LHC and other future colliders. The LHC in its high luminosity version could probe these couplings through searches of rare decays.% and single production of top quark. 
Being a top factory with very large statistics, the sensitivity of HL-LHC is quite competitive compared to any other planned future collider. However, when it comes to disentangling the effects of many possible couplings (like $tq\gamma$, $tqZ$, $tqg$), colliders with electron in its initial states stand with clear advantages. In this work we have demonstrated the uniqueness of high energy electron-proton collider in this respect.

The process studied, namely the production of single top quark in association with a light jet and scattered electron in electron-proton collisions ($e^-p\rightarrow e^-tj$) in the planned high energy facility of FCC-he, has the potential to fingerprint the presence of top quark FCNCs involving photon and gluon. We have demonstrated the sensitivity of the collider with electron beam of 60 GeV and proton beam of 50 TeV. In addition, the possibility of electron beam polarizations along with the consideration of suitable angular asymmetries help us to distinguish the different types of couplings. We have considered the leptonic decay of the top quark in our study. A way to fully reconstruct the top quark, even with a missing neutrino as presented in \cref{ch_5:spin_matrix} is made use of in reconstructing the events. The process has no SM background with the same parton level final state. However all possible processes that could mimic the final state at the detector are considered in the study. 
We could contain the background events to the level of a few tens at the expected luminosity of 2 ab$^{-1}$. With such reduced background, we have demonstrated that the presence of couplings to the level of a few times 10$^{-2}$ TeV$^{-1}$ can be probed at this collider with the process considered.  This corresponds to the branching fractions of ${\rm BR}(t\to u\gamma)\le 4 - 7 \times 10^{-6}$ and ${\rm BR}(t\to c\gamma) \le 1-2 \times 10^{-5}$, depending on if the coupling is right-handed or left-handed. The corresponding limits on the gluon couplings lead to  ${\rm BR}(t\to ug)\le 1.7 \times 10^{-5}$ and ${\rm BR}(t\to cg) \le 3-4 \times 10^{-5}$.
 The distinction in the sensitivity of the right-handed and left-handed couplings are mainly due to their effect on the angular distribution of the scattered electron as well as on the top polarisation, which affect the signal selection and thus the fiducial cross section.
Further, effective use of top quark polarization asymmetries, forward-backward asymmetry of the scattered electron, along with the effect of electron beam polarization on the cross section, it is possible to fully disentangle the effects of the four relevant couplings considered in the study.

\begin{acknowledgments}
	We acknowledge fruitful discussions within the Higgs and top physics study for LHeC and FCC-he group in customising the detector card used in this analysis. SB and PP thank the DST-FIST grant SR/FST/PSIl-020/2009 for offering the computing resources needed by this work. PP is thankful to Department of Physics, Concordian University, for their hospitality during the manuscript preparation of this work.
\end{acknowledgments}
%
%----------------------------------------------------------------------------------------
%	REFERENCE LIST
%----------------------------------------------------------------------------------------

%\nocite{*}

\bibliography{biblio}% Produces the bibliography via BibTeX.

\end{document}